
\input phyzzx

\def\np{Nucl. Phys.}
\def\pl{Phys. Lett.}

\def\cmp{Comm. Math. Phys.}
\def\ijmp{Int. J. Mod. Phys.}

\def\ex{{\hbox{\rm e}}}
\def\azb{A_{\bar z}}
\def\az{A_z}
\def\im{{\hbox{\rm Im}}}
\def\tr{{\hbox{\rm Tr}}}

\def\ex{{\hbox{\rm e}}}
\def\azb{A_{\bar z}}
\def\az{A_z}
\def\bzb{B_{\bar z}}
\def\bz{B_z}
\def\czb{C_{\bar z}}
\def\cz{C_z}
\def\dzb{D_{\bar z}}
\def\dz{D_z}

\def\im{{\hbox{\rm Im}}}
\def\mod{{\hbox{\rm mod}}}
\def\tr{{\hbox{\rm Tr}}}

\tolerance=500000
\overfullrule=0pt
\Pubnum={US-FT-15/91}
\date={December, 1991}
\pubtype={}
\titlepage

\title{COSET CONSTRUCTIONS IN CHERN-SIMONS GAUGE THEORY}
\author{J. M. Isidro, J. M. F. Labastida and A. V. Ramallo}
\address{Departamento de F\'\i sica de Part\'\i culas \break
Universidad de Santiago \break
E-15706 Santiago de Compostela, Spain}

\abstract{Coset constructions in the framework of Chern-Simons topological
gauge theories are studied. Two examples are considered: models of the types
${U(1)_p\times U(1)_q\over U(1)_{p+q}}\cong U(1)_{pq(p+q)}$ with $p$ and $q$
coprime integers, and  ${SU(2)_m\times SU(2)_1\over SU(2)_{m+1}}$. In the
latter
case it is shown that the Chern-Simons wave functionals can be identified with
t
   he
characters of the minimal unitary models, and  an  explicit representation of
the knot (Verlinde) operators acting on the space of  $c<1$ characters is
obtained.}

\endpage
\pagenumber=1
\sequentialequations

{\it 1. Introduction. }Three years ago Witten  found a remarkable
equivalence between the Hilbert space of three-dimensional (3D) topological
Chern-Simons (CS) theory and the space of conformal blocks of the
two-dimensional (2D) Wess-Zumino-Witten (WZW) model with the same gauge group
and level \REF\wit{E. Witten \journal\cmp&121(89)351}
[\wit]. This equivalence has been analyzed and
made more explicit in subsequent works by several authors
\REF\mszoo{G. Moore and N. Seiberg \journal\pl&B220(89)422}
\REF\emss{S. Elitzur, G. Moore, A. Schwimmer and N. Seiberg
\journal\np&B326(89)108}
\REF\bn{M. Bos and V.P. Nair \journal\pl&B223(89)61
\journal\ijmp&A5(90)959}
\REF\nos{J.M.F. Labastida and A.V. Ramallo \journal\pl&B227(89)92
\journal\pl&228(89)214,
{\sl Nucl. Phys.} {\bf B} (Proc. Suppl.) {\bf 16B} (1990) 594}
\REF\knot{J.M.F. Labastida, P.M. Llatas and A.V. Ramallo
\journal\np&B348(91)651}
[\mszoo,\emss,\bn,\nos,\knot]. An operator formalism was developed in
[\nos ,\knot] where it was shown that  the  CS wave functionals can be
identified with the characters of the corresponding WZW model. One may ask
whether it is essential to have a current algebra in the 2D Rational Conformal
Field Theory (RCFT) in order to describe its conformal blocks as wave
functionals of a 3D CS gauge theory. As conjectured in [\mszoo], Witten's
equivalence between the two theories could be extended to any RCFT by using GKO
constructions
\REF\GKO{P. Goddard, A. Kent and D. Olive
\journal\pl&B152(85)88 \journal\cmp&103(86)535} [\GKO]
and, therefore, the existence of a current algebra would not be required in
general. In this paper we  verify this conjecture by showing
how the operator formalism constructed in  [\nos,\knot] can be
used to describe coset models. Two examples will be considered. First, we
analyze an abelian coset model which will help us to find the good quantization
variables for  CS coset constructions. Secondly we study the non-abelian
${SU(2)_m\times SU(2)_1\over SU(2)_{m+1}}$ models which are well-known to be
equivalent to the minimal unitary models with central charge less than
unity [\GKO]. We will be able to formulate in this case a quantum-mechanical
problem having the $c<1$ characters as states and the Verlinde operators
\REF\verope{E. Verlinde \journal\np&B300(88)360}[\verope]
associated to any toral knot as observables. In the present letter we give a
brief report of our results. An extended version containing details and
extensions of our calculations will be presented elsewhere.

{\it 2. Abelian Coset Models. } Let us study the quantization of an abelian
topological Chern-Simons (CS) theory defined on a three manifold $M$
with action,
$$
S={p\over
2\pi}\int_M A\wedge dA
+{q\over
2\pi}\int_M B\wedge dB,
\eqn\uno
$$
where $p$ and $q$ are two coprime  integers and $A$ and $B$ are $U(1)$
one-form connections. We shall restrict ourselves to manifolds $M$ that
can be put at least locally as $\Sigma\times R ^1$ where $\Sigma$
is  a compact Riemann surface without boundary, and  $R ^1$ will be taken as
time direction. On $\Sigma$ we introduce
a system of local complex coordinates $z$ and $\bar z$. The corresponding
components of the gauge fields parallel to $\Sigma$ will be denoted as
$A_z$, $B_z$ and $A_{\bar z}$, $B_{\bar z}$.
In order to proceed to a canonical
quantization of the theory let us choose the gauge in which the components of
$A$ and $B$ along $R^1$ vanish. In this gauge the holomorphic components of
the gauge fields are canonically conjugate to the antiholomorphic ones in
such a way that the only non-vanishing commutation relations on the equal-time
surface $\Sigma$ are,
$$
[\az(\sigma),\azb(\sigma')]= -{\pi \over 2p} \delta^2(\sigma-\sigma'),
\,\,\,\,\,\,\,\,\,
[\bz(\sigma),\bzb(\sigma')]= -{\pi \over 2q} \delta^2(\sigma-\sigma'),
\eqn\dos
$$
where $\sigma^i$ represent local real coordinates on $\Sigma$.
When the manifold $M$ has no boundary the action \uno\ is invariant under
gauge transformations: $A\rightarrow A +g^{-1}dg$,
$B\rightarrow B +h^{-1}dh$,
with $g=e^\alpha$ and $h=e^\beta$ . In our
gauge these transformations are generated by the operator
$$
Q(\alpha, \beta)= - {2\over \pi}\int_\Sigma(\alpha pF_{z{\bar z}}[A]
+\beta qF_{z{\bar z}}[B] ),
\eqn\tres
$$
where $F_{z{\bar z}}[A]$ and $F_{z{\bar z}}[B]$ are  field strength
tensors (\ie, $ F_{z{\bar z}}[A]= \partial _z \azb - \partial _{\bar z} \az
$, etc).
The operator that implements gauge transformations in the
Hilbert space  is simply obtained by exponentiating the generator \tres:
$G(\alpha,\beta)=\ex^{Q(\alpha,\beta)}$. The physical states are those left
invariant by the action of $G(\alpha,\beta)$. This is the Gauss law, which
is enough to determine the states of the theory. From the commutation
relations \dos\ we conclude that one can represent these states as wave
functionals depending on the antiholomorphic components of the gauge fields:
$\Psi[\azb,\bzb]$. As $A$ and $B$ are independent fields, the solution to
the Gauss law, $G(\alpha,\beta)\Psi=\Psi$, can be factorized as,
$$
\Psi[\azb,\bzb]=\Phi_p[\azb]\Phi_q[\bzb].
\eqn\cuatro
$$
Using the commutation relations \dos\ and  the explicit form of the
generator \tres, the Gauss law implies the following equation for
$\Phi_p[\azb]$:
$$
\Phi_p[\azb+\partial_{\bar z} \alpha]=
\ex^{\big (-{p\over \pi}\int_\Sigma \partial_z \alpha \partial_{\bar z}\alpha
-{2p\over \pi} \int_\Sigma \partial_z \alpha \azb \big)}
\Phi_p[\azb].
\eqn\cinco
$$
($\Phi_q[\bzb]$ satisfies a similar equation). In order to solve \cinco , let
us restrict ourselves to the case in which $\Sigma$ is a torus $T^2$ with
modular parameter $\tau$. The holomorphic one-form $\omega(z)$ is defined by
means of its integrals along the cycles  ${\bf A}$ and  ${\bf B}$ of the
canonical homology basis: $\int_{\bf A}\omega =1$, $\int_{\bf B}\omega =\tau$.
Our conventions are such that the ${\bf A}$-cycle is contractible in the solid
torus and $\int_{T^2} d^2\sigma\,\omega (z){\overline{\omega (z)}}= \tau_2$,
with $\tau_2=\im \tau$. A general gauge transformation
$g: T^2\rightarrow U(1)$ can be parametrized as
$g(z)=\exp[\alpha(z)+\alpha_{mn}(z)]$, where $\alpha(z)$ is a single-valued
function on $T^2$, whereas $e^{\alpha_{mn}}$ is a function that winds
$m$ ($n$) times around the $U(1)$ group as we move along the ${\bf A}$
(${\bf B}$) cycle on $T^2$. These large gauge transformations can be
parametrized in general with the help of the holomorphic one-forms,
$$
\alpha_{mn}= -{\pi\over \tau_2}(n+m{\bar
\tau})\int^z\omega(z) +
{\pi\over\tau_2}(n+\tau m)\int^{\bar z}{\overline{\omega(z)}}.
\eqn\seis
$$
We are now ready to obtain the general solution to the Gauss law. Let us
first  parametrize $\azb$ and $\az$ as
$\azb= \partial_{\bar z} \chi^A +{\pi{\bar \omega}\over \tau_2}a$,
$\az=-\azb^\dagger$ where $\chi^A$ is a single-valued function and $a$ is a
constant complex number containing all the information about the holonomy of
the gauge field around the non-trivial cycles of $T^2$. Under the gauge
transformation induced by $g$, the variables $\chi^A$ and $a$ change as
follows:  $\chi^A \rightarrow \chi^A + \alpha$,
$a \rightarrow a +n+\tau m $. The general solution to the Gauss law \cinco\
can be easily expressed in terms of $\chi^A$ and $a$. It is a linear
combination of the functions:
$$
\Phi_{j,p}[\azb]=\ex^{-{p\over \pi}\int_{T^2}\partial_z\chi^A
\partial_{\bar z}\chi^A}\varphi_{j,p}(a,\tau),
\eqn\siete
$$
where $j$ is an integer and $\varphi_{j,p}(a,\tau)$ is given by
$$
\varphi_{j,p}(a,\tau)=[\eta (\tau)]^{-1}
\ex^{\pi a^2\tau_2^{-1}p}
\theta\big [^{{j\over 2p}}_0\big](2pa|2p\tau).
\eqn\ocho
$$
In \ocho\  $\eta(\tau)$ is the Dedekind $\eta$-function, and
$\theta\big [^a_b\big ](z|\tau)$ denotes the Jacobi theta function with
characteristics
\REF\mun{See, for example, D. Mumford, {\it Tata Lectures on Theta},
Birkh\"auser, Basel, 1983}
[\mun]. Notice that $\varphi_{j+2p,p}=\varphi_{j,p}$, and therefore
there are only $2p$ independent solutions to the Gauss law equation \cinco.
Obviously, the general form of $\Phi_q[\bzb]$  can be obtained in the
same way. If we parametrize $\bzb$ as $\bzb= \partial_{\bar z} \chi^B +
{\pi{\bar \omega}\over \tau_2}b$, the $2q$ independent solutions
$\Phi_{k,q}[\bzb]$ are given by eqs. \siete\ and \ocho\ with the changes
$j\rightarrow k$, $p\rightarrow q$,  $\chi^A\rightarrow \chi^B$ and
$a\rightarrow
b$. From \siete\ and \ocho\ it is evident that our states can be identified
with the characters of a RCFT with an abelian
Kac-Moody symmetry. In the case at hand the basis of our Hilbert space are
the characters of a  $U(1)_p \times U(1)_q$ current algebra. This is nothing
but
a particular case of  Witten's correspondence between the Hilbert space of CS
theories and the space of conformal blocks of RCFT's.

In a path integral approach it is possible to give a representation to the
solutions of the Gauss law as functional integrals over gauge field
configurations. Suppose that our three-dimensional manifold is a solid torus
whose boundary is the torus $T^2$ on which our wave functionals are defined.
We can represent $\Psi[\azb,\bzb]$ as,
$$
\Psi [\azb,\bzb]=\int [DA DB] W_{j,p}^{\gamma}(A) W_{k,q}^{\gamma '}(B)
\exp\big [iS-{1\over\pi}\int_{T^2} d^2\sigma (p\az\azb +q\bz\bzb) \big ].
\eqn\nueve
$$
In \nueve\ we integrate over gauge field configurations on the solid torus
having fixed values of $\azb$ and $\bzb$  on the
boundary $T^2$. We have also inserted in our functional integral Wilson line
operators associated to closed curves $\gamma$ and $\gamma'$ in
the solid torus. $W_{j,p}^{\gamma}(A)$ is given by the equation
$$
W_{j,p}^{\gamma}(A)=\exp[-j\int_{\gamma}\,A].
\eqn\diez
$$
Our Gauss law selects flat connections, which means that the path integral
\nueve\ only depends on the homotopy class of the curves $\gamma$ and
$\gamma'$. If in addition the $U(1)$ charges $j$ and $k$ are integers,
then $W_{j,p}^{\gamma}(A)$ and $W_{k,q}^{\gamma'}(B)$ are gauge invariant under
large gauge transformations that wind around $U(1)$ as one goes along
$\gamma$ and $\gamma'$. Therefore these operators are the natural
observables for a topological gauge theory.
The functional integral representation \nueve\ implies the following inner
product for wave functionals:
$$
({\tilde\Psi},\Psi)=\int [D\az D\bz D\azb D\bzb]\exp\big [
{2\over \pi}\int_{T^2} d^2\sigma(p\az\azb+q\bz\bzb\big]
{\overline{{\tilde \Psi}(\azb,\bzb)}}\Psi(\azb,\bzb).
\eqn\once
$$
It can be checked [\nos,\knot] that, with the inner product \once, the states
$\Psi_{j,k}= \Phi_{j,p}\Phi_{k,q}$ are orthonormal. Moreover, the elements of
this basis of our Hilbert space can be obtained by means of very specific
insertions in the path integral \nueve: the state $\Psi_{j,k}$ is created on
$T^2$ by  Wilson lines having $U(1)_p$ and $U(1)_q$ charges $j$ and $k$ for
curves homotopic to the ${\bf B}$ cycle of $T^2$. We get in this way a
correspondence between states and observables similar to what happens in RCFT.
In order to check this fact let us consider Wilson line operators associated to
curves lying completely on $T^2$. These curves are the so-called toral knots,
and they are characterized  by two coprime integers $r$ and $s$ representing
the number of times the path of the knot winds around the ${\bf A}$ and ${\bf
B}$ cycles respectively. As we noticed above, the effect of inserting one of
these operators only depends on the homotopy class in the solid torus of the
corresponding path. Of all the possible curves $\gamma$ and $\gamma'$ in
\nueve,  the toral knots are the most interesting ones in an operator
formalism because they can be represented by well-defined operators acting on
the Hilbert space of  CS wave functionals. Notice first of all that,  with the
parametrization of the gauge fields $A$ and $B$ on $T^2$ given above, the
Wilson lines for closed curves on the torus only depend on the holonomy
parameters $a$ and $b$ (and on their complex conjugates $\bar a$ and $\bar b$
appearing in the parametrization of $\az$ and $\bz$). From the canonical
commutation relations \dos\ (or equivalently from the inner product \once) we
can read off the following commutators:   $$
[{\bar a},a]= {\tau_2\over 2\pi p},
\,\,\,\,\,\,\,\,\,\,
[{\bar b},b]= {\tau_2\over 2\pi q},
\eqn\doce
$$
and therefore $\bar a$ and $\bar b$ can be represented as,
$$
{\bar a}={\tau_2\over 2\pi p}{\partial\over\partial a},
\,\,\,\,\,\,\,\,\,\,
{\bar b}={\tau_2\over 2\pi q}{\partial\over\partial b}.
\eqn\trece
$$
It is now immediate to write down the operator representation of \diez\ for
$\gamma$ equal to the toral knot $(r,s)$ (\ie,  for $\gamma=r{\bf A}+s{\bf
B}$),
$$
W_{j,p}^{(r,s)}(A)=\exp{ \big[{-j\pi \over \tau_2}(r+s{\bar\tau})a
+{j\over 2p}(r+s\tau){\partial\over\partial a}\big]}.
\eqn\catorce
$$
Using well-known properties of the theta functions it is now easily proved
[\nos] that $W_{l,p}^{(0,1)}(A)W_{m,q}^{(0,1)}(B)\Psi_{j,k}=\Psi_{j+l,k+m}$,
which implies that Wilson line operators for ${\bf B}$ cycles create
$U(1)_p\times U(1)_q$ charges. In view of this result it is natural to
associate the ``vacuum state" $\Psi_{0,0}$ (which has vanishing charges) to the
path integral  \once\ with no Wilson lines inserted. All the independent states
of the Hilbert space are obtained by acting on the vacuum with the ``creation"
operators  $W_{l,p}^{(0,1)}(A)$ and $W_{m,q}^{(0,1)}(B)$. These operators are
nothing but the Verlinde operators [\verope], which represent the primary
fields of the current algebra on the space of characters. In fact for any $r,s$
in \catorce\ we get a representation of the Verlinde algebra (fusion rules) for
the underlying current algebra:
$W_{j,p}^{(r,s)}(A)W_{l,p}^{(r,s)}(A)=W_{j+l,p}^{(r,s)}(A)$, and similarly for
the $B$ field.

So far we have described the quantization of a topological gauge theory
consisting of two independent Chern-Simons fields $A$ and $B$. We will now
show how this system can also be described in terms of new gauge
fields that will allow us to perform a coset construction. First of all let us
define new one-form connections,
$$
C={pA+qB\over {p+q}},
\,\,\,\,\,\,\,\,\,\,\,\,\,\,\,
D={A-B\over{p+q}},
\eqn\quince
$$
whose inverse relations are:
$$
A=C+qD,
\,\,\,\,\,\,\,\,\,\,\,\,\,\,\,
B=C-pD.
\eqn\dieciseis
$$
We will show below that the quantum mechanical problem of quantizing the
$U(1)_p\times U(1)_q$ CS theory given by the action \uno\ naturally gives rise
to the introduction of the gauge connections $C$ and $D$ defined in \quince.
In fact we shall prove that the action, observables and states of the theory
\uno\ can be decomposed in terms of a $U(1)_{p+q}\times U(1)_{pq(p+q)}$ CS
theory for the gauge fields $C$ and $D$. First of all notice that, in the gauge
in which the time components of the fields $A$ and $B$ vanish, the commutation
relations \dos, together with \quince, imply:
$$
[\cz(\sigma),\czb(\sigma')]= -{\pi \over {2(p+q)}} \delta^2(\sigma-\sigma'),
\,\,\,\,\,\,\,\,\,\,\,\,\,
[\dz(\sigma),\dzb(\sigma')]= -{\pi \over {2pq(p+q)}} \delta^2(\sigma-\sigma').
\eqn\diecisiete
$$
Moreover $C$ and $D$ commute and therefore they can be considered as
independent fields. Equation \diecisiete\ seems to indicate that $C$ and $D$
describe $U(1)$ theories with levels $p+q$ and $pq(p+q)$ respectively. This is
indeed the case, as can  be seen from the form that the action $S$
given in \uno\ takes in these variables,
$$
S={(p+q)\over
2\pi}\int_M C\wedge dC+{pq(p+q)\over2\pi}\int_M D\wedge dD.
\eqn\dieciocho
$$
In fact, the same decomposition also holds for the observables of the theory.
Using  equation \dieciseis\ and the canonical commutation relations
\diecisiete, it is easy to relate Wilson lines in both sets of variables for a
given closed curve $\gamma$,
$$
W_{j,p}^{\gamma}(A) W_{k,q}^{\gamma }(B)=
W_{j+k,p+q}^{\gamma}(C) W_{jq-kp,pq(p+q)}^{\gamma }(D).
\eqn\diecinueve
$$
Notice that the Wilson lines in the variables $C$ and $D$ resulting
from the product of two gauge invariant operators
$W_{j,p}^{\gamma}(A) W_{k,q}^{\gamma }(B)$ in \diecinueve\ have integer
charges and therefore are also invariant under large gauge transformations of
the fields $C$ and $D$. They are thus good observables for the theory defined
by the action \dieciocho\ (\ie, when $C$ and $D$ are considered as
fundamental fields).

The Hilbert space of states of the $U(1)_p\times U(1)_q$ theory can also be
decomposed in terms of $U(1)_{p+q}\times U(1)_{pq(p+q)}$ CS wave functionals.
The following relation holds:
$$
\Phi_{j,p}[\azb]\Phi_{k,q}[\bzb]=
\sum_{l\in Z_{p+q}}\Phi_{j+k+2pl,p+q}[\czb]\Phi_{jq-kp+2pql,pq(p+q)}[\dzb].
\eqn\veinte
$$
In \veinte\ $\Phi_{i,p+q}[\czb]$ and $\Phi_{i,pq(p+q)}[\dzb]$ are given by
equation \siete\ with the appropriate changes in the arguments. If $p$ and $q$
are coprimes all the $2(p+q)$ ($2pq(p+q)$) states $\Phi_{i,p+q}[\czb]$
($\Phi_{i,pq(p+q)}[\dzb]$ ) appear on the right-hand side of
\veinte\ when the different $U(1)_p\times U(1)_q$ states are decomposed. In
order to prove \veinte\ it is enough to check it for the case $j=k=0$. The
proof for the other cases follows from the Wilson line decomposition law
\diecinueve\ and from the fact that in CS theory the Wilson line operators
for the ${\bf B}$ homology cycle create  the corresponding character when
acting
on the vacuum. For $j=k=0$ in \veinte\ we need to study the following product:
$$
\theta(2pa|2p\tau)\theta(2qb|2q\tau)=
\sum_{m\in Z}\sum_{n\in Z}\ex^{i\pi\tau(2pm^2+2qn^2)}
\ex^{2\pi i(2pma+2qnb)},
\eqn\veintiuno
$$
where we have denoted by $\theta(z|\tau)$ the theta functions with vanishing
characteristics. Suppose  we parametrize $C$ and $D$ on $T^2$ in a
way similar to what we did for $A$ and $B$ and let us denote by $c$ and $d$
the holonomy parameters for $C$ and $D$ respectively. The right-hand side of
\veintiuno\ can be rewritten as a sum of theta functions with $c$ and $d$ in
their arguments if one introduces the new summation indices $r$ and $s$,
$$
m-n=(p+q)s,
\,\,\,\,\,\,\,\,\,\,\,\,
pm+qn=(p+q)r.
\eqn\veintidos
$$
Notice that in general $r$ and $s$ are not integers. Let us obtain their
general form.
{}From \veintidos\ follows that  $m=r+qs$, $n=r-ps$, and therefore $r$ and
$s$ satisfy the constraints $r+qs\in Z$ and $r-ps\in Z$. In order to solve
these
constraints let us first parametrize $r$ and $s$ in the form dictated by
\veintidos,
$$
r={\bar r}+{k\over p+q},
\,\,\,\,\,\,\,\,\,\,\,\,
s={\bar s}+{l\over p+q}.
\eqn\veintitres
$$
Now $\bar r$ and $\bar s$ are independent integers and $k,l\in Z_{p+q}$. The
constraints on $r$ and $s$ translate into the simple condition $k=pl\,\,\,
\mod \,(p+q)$, which solves for $k$ in terms of $l$. The sums in $\bar r$ and
$\bar s$  build theta functions in the variables $c$ and $d$ respectively. The
final result is,
$$
\eqalign{&\theta(2pa|2p\tau)\theta(2qb|2q\tau)=\cr
&\sum_{l\in Z_{p+q}}
\theta \Bigg[
{{2pl\over 2(p+q)}\atop 0}\Bigg](2(p+q)c|2(p+q)\tau)
\theta \Bigg[
{{2pql\over 2pq(p+q)}\atop 0}\Bigg](2pq(p+q)d|2pq(p+q)\tau)\cr}_{,}
\eqn\veinticuatro
$$
which is the crucial equation needed to prove \veinte\ for $j=k=0$.

{}From eqs. \dieciocho, \diecinueve\ and \veinte\ we conclude that, as
indicated above, $C$ and $D$ are the natural variables to decompose a
$U(1)_p\times U(1)_q$ CS theory in terms of states with well-defined quantum
numbers under $U(1)_{p+q}\times U(1)_{pq(p+q)}$. Plugging this decomposition
into our Hilbert space inner product \once\ and integrating out the $C$ field,
we end up with an effective quantum-mechanical problem for the remaining field
$D$. Using \dieciseis\ we obtain that the measure for the inner product \once\
is $\exp\big [{2\over \pi}\int d^2\sigma((p+q)\cz\czb+pq(p+q)\dz\dzb )\big]$.
We
thus see that the effective problem in $D$ corresponds to an abelian CS theory
with level equal to $pq(p+q)$. The integration of the $C$ field is the CS
version of the standard coset procedure of modding out the degrees of
freedom corresponding to $C$. In the case at hand we are
performing the coset ${U(1)_p\times U(1)_q\over U(1)_{p+q}}$. Eqs. \dieciocho
-\veinte\ mean that,
$$
{U(1)_p\times U(1)_q \over U(1)_{p+q}}\cong U(1)_{pq(p+q)},
\eqn\veinticinco
$$
which is a well-known result at the level of two-dimensional current
algebras\foot{If $p$ and $q$ are not coprime, then the level of the $U(1)$
coset
algebra is ${pq(p+q)\over (p,q)^2}$ where $(p,q)$ is the greatest common
divisor of $p$ and $q$. In this case the gauge fields $C$ and $D$ must be
defined as in \quince\ with the changes $p\rightarrow {p\over(p,q)}$,
$q\rightarrow {q\over(p,q)}$.}[\mszoo].

In order to understand the implications of the change of variables discussed
above, let us remark that according to eq. \veinte\ every $U(1)_p\times
U(1)_q$ state splits into a sum of $p+q$ states of $\,\,$ $U(1)_{p+q}\times
U(1)_{pq(p+q)}$. Such a proliferation of states in the new variables
originates from the fact that both types of wave functionals  are the solutions
to different Gauss laws. When the gauge quantization problem is formulated in
terms of $C$ and $D$ as fundamental fields, one expects that the operators
$W_{l,p+q}^{\gamma}(C)$ and  $W_{n,pq(p+q)}^{\gamma }(D)$ be gauge invariant
for any integer charges $l$ and $n$ and any closed curve $\gamma$. Suppose that
we perform abelian gauge transformations in $A$ and $B$: $A\rightarrow A
+g^{-1}dg$, $B\rightarrow B +h^{-1}dh$. Let us denote by $n_g$ and $n_h$ the
winding numbers of $g$ and $h$ along $\gamma$:
$$
n_g={1\over 2\pi i}\int_{\gamma}g^{-1}dg,
\,\,\,\,\,\,\,\,\,\,\,\,\,\,\,
n_h={1\over 2\pi i}\int_{\gamma}h^{-1}dh.
\eqn\veintiseis
$$
If we require gauge invariance of $W_{l,p+q}^{\gamma}(C)$ and
$W_{n,pq(p+q)}^{\gamma }(D)$ for any $l,n\in Z$, we must restrict ourselves to
gauge transformations such that certain combinations of $n_g$ and $n_h$ are
integers (see eq. \quince):
$$
{pn_g+qn_h\over p+q}=r,
\,\,\,\,\,\,\,\,\,\,\,\,\,\,
{n_g-n_h\over p+q}=s,
\,\,\,\,\,\,\,\,\,\,\,\,\,\,
r,s\in Z.
\eqn\veintisiete
$$
{}From eq. \veintisiete\ one gets $n_g=r+qs$, $n_h=r-ps$, which in particular
imply that $n_g-n_h=(p+q)s$. Therefore $n_g$ and $n_h$ must be equal modulo
$p+q$,
$$
n_g=n_h\,\,\,\,\,\mod\,\,(p+q).
\eqn\veintiocho
$$
Remember that when we take Wilson lines for
$A$ and $B$ with integer charges  as basic observables of our theory, $n_g$ and
   $n_h$ can be any two independent
integers. Therefore we see that when we perform the quantization of the system
taking $C$ and $D$ as fundamental fields, we are actually reducing the global
gauge symmetry of the physical states and thus the Gauss law becomes less
restrictive. This explains why we get more gauge-invariant states in the new
variables.

We may ask how one can modify the path integral \nueve\ in such a way that a
state with well-defined $U(1)_{p+q}\times U(1)_{pq(p+q)}$ quantum numbers is
obtained. In \nueve\ the path integral is performed over $A$ and $B$ and we
wish to obtain a state of those created by acting on the vacuum
$\Phi_{0,p+q}(\czb)\Phi_{0,pq(p+q)}(\dzb)$ with ${\bf B}$-cycle Wilson line
operators for the gauge fields $C$ and $D$. This situation is reminiscent of
the orbifold constructions
\REF\Gins{ See, for example, P. Ginsparg, ``Applied Conformal Field Theory", in
{\it Fields, Strings and Critical Phenomena}, ed. by E. Brezin and J.
Zinn-Justin, North Holland, Amsterdam, 1990}
[\Gins] in which one describes a system (for example the Ashkin-Teller
model) in terms of fields (a scalar field in this example) which are not
the basic variables of the system (the spin fields). In order to produce the
correct spectrum of these models one has to introduce a
projector in the path integral which is equivalent to sum over all possible
twisted sectors of the theory. In our case let us show that something similar
happens. We first notice that eq. \veinte\ can be inverted. The products of
theta-functions
 appearing in the right-hand side of \veinte\ can be written as,
$$
\eqalign{&\varphi_{j+k,p+q}(c,\tau)\varphi_{jq-kp,pq(p+q)}(d,\tau)=\cr
&={1\over p+q}\sum_{l\in Z_{p+q}}\ex^{-{i\pi l(jq-kp)\over pq(p+q)}}
{\ex^{{\pi p a^2\over \tau_2}}\over \eta(\tau )}
\theta \Bigg[{{j\over 2p}\atop {l\over p+q}}\Bigg](2pa|2p\tau)
{\ex^{{\pi q b^2\over \tau_2}}\over \eta(\tau )}
\theta \Bigg[{{k\over 2q}\atop {-l\over p+q}}\Bigg](2qb|2q\tau). \cr}
\eqn\veintinueve
$$
An important point in analyzing this equation is the fact that the lower
characteristics appearing in \veintinueve\ can be generated by acting with
Wilson line operators for the ${\bf A}$-cycle. Suppose  we consider the case
$j=k=0$ (using \diecinueve\ for a ${\bf B}$-cycle the analysis can be trivially
extended to any $j$ and $k$). From the operator representation of the Wilson
lines (see eq. \catorce) and the decomposition law \diecinueve\ one gets a
relation between both types of vacua,
$$
\Phi_{0,p+q}(\czb)\Phi_{0,pq(p+q)}(\dzb)=
\Big ({1\over{p+q}}\sum_{l\in Z_{p+q}} W_{l,pq(p+q)}^{(1,0) }(D) \Big )
\Phi_{0,p}(\azb)\Phi_{0,q}(\bzb).
\eqn\treinta
$$
Let us now define the operator $P$ as follows:
$$
P={1\over{p+q}}\sum_{l\in Z_{p+q}} W_{l,pq(p+q)}^{(1,0) }(D).
\eqn\treintayuno
$$
It is easy to check that $P$ is a projector when acting on the vacuum
$\Phi_{0,p}(\azb)\Phi_{0,q}(\bzb)$  (\ie, $P^2=P$ and
$P^\dagger=P$). Finally eq. \treinta\ allows us to represent the
$U(1)_{p+q}\times U(1)_{pq(p+q)}$ vacuum as a path integral over the fields $A$
and $B$,
$$
\Phi_{0,p+q}(\czb)\Phi_{0,pq(p+q)}(\dzb)=\int [DA DB] P
\exp\big [iS-{1\over\pi}\int_{T^2} d^2\sigma (p\az\azb +q\bz\bzb) \big ].
\eqn\treintaydos
$$
The extension of this result for arbitrary charges $j$ and $k$ is
straightforward provided the adequate Wilson line operators are introduced in
the functional integral.

{\it 3. Non-Abelian Coset Models.}
Let us now extend the previous formalism to a non-abelian case. Our aim is
to construct the CS coset ${SU(2)_m\times SU(2)_1\over SU(2)_{m+1}}$ which
gives
rise [\GKO] at the two-dimensional level to the  minimal unitary
representations of the Virasoro algebra with central charge $c_m=1-{6\over
(m+2)(m+3)}$, for $m=1,2,...$ [\Gins]. Our starting point will be the
non-abelian CS theory for two independent gauge fields $A$ and $B$ having
levels $m$ and $1$ respectively,
$$
S={m\over4\pi}\int_M Tr \Big [ A\wedge dA +{2\over 3}A\wedge A\wedge A \Big ] +
{1\over4\pi}\int_M Tr \Big [ B\wedge dB +{2\over 3}B\wedge B\wedge B \Big ].
\eqn\treintaytres
$$
As was proved in [\nos ,\knot], the operator formalism of a non-abelian CS
theory on $T^2$ gives rise to an effective quantum-mechanical problem in the
holonomy part of the gauge connection, which takes values in the Cartan
subalgebra of the gauge group.  For an  $SU(2)$ gauge field $A$ on the torus,
 after taking the gauge $A_0=0$, this part of the connection can be
parametrized as $\azb={\pi a\over 2 \tau_2}{\bar\omega}\sigma_3$, and
$\az=-\azb^\dagger$,  where $\sigma_3$ is a Pauli matrix and we are in the
gauge
$A_0=0$. The states that solve the effective Gauss law are of the form
[\nos,\knot], $$ \Phi_{j,m}={\lambda_{j,m+2}(a)\over \Pi (a)},
\eqn\treintaycuatro
$$
being
$$
\lambda_{j,m+2}(a)=\ex^{\pi(m+2)a^2\over 4\tau_2}
\Big [\Theta_{j+1,m+2}(a,\tau,0)-\Theta_{-j-1,m+2}(a,\tau,0) \Big ],
\eqn\treintaycinco
$$
where $\Theta_{j,k}$ are classical theta-functions
\REF\Kac{V.G.Kac, Infinite Dimensional Lie Algebras,  (Birkh\"auser,
Basel, 1983)}
 [\mun,\Kac] and
$\Pi(a)=\lambda_{0,2}(a)$. The wave functionals $\Phi_{j,m}(a)$ represent the
characters of an $SU(2)$ Kac-Moody algebra at level $m$ for an isospin $j/2$
\REF\GW{D. Gepner and E. Witten \journal\np&B278(86)493}[\Kac,\GW ].
Using the periodicity properties of the theta-functions one can check that
there
are $m+1$ independent states labeled by $j=0,\cdots,m$. The wave functionals
\treintaycuatro\ are symmetric under the Weyl reflection $a\rightarrow -a$,
although both the numerator and the denominator in \treintaycuatro\ are
antisymmetric. The inner product measure in the effective Hilbert space is
${da d{\bar a}\over 2\sqrt{\tau_2}}\Pi (a) {\overline \Pi(a)}
\ex^{{-(m+2)\over 2\tau_2}a{\bar a}}$. As the $\Pi$ factors coming from the
meas
   ure
and the states cancel in the inner product, we can ignore them everywhere and
take the numerator of \treintaycuatro\ as wave function . In this basis of
states the operator formalism greatly simplifies [\nos,\knot] since the basic
commutator is just  $[{\bar a},a]={2\tau_2\over \pi(m+2)}$. The observables of
the theory are Wilson lines associated to representations of $SU(2)$,
$W_{j,m}^\gamma(A)= \tr_j \exp[-\int_{\gamma}A]$, the trace taken in the
representation of isospin $j/2$ of $SU(2)$. Let us denote by $\Lambda_j$ the
set of weights of the $SU(2)$ irreducible representation of isospin ${j\over
2}$ (\ie, the eigenvalues of the generator of the Cartan subalgebra). In our
conventions  $\Lambda_j=\{-j,-j+2,\cdots ,j-2,j\}$. The $SU(2)$ operators
$W_{j,m}^\gamma(A)$ can be written as a sum of  abelian Wilson lines whose
charges are the set of weights $\Lambda_j$. For a toral knot $\gamma=r{\bf
A}+s{\bf B}$ one gets the operator
$W_{j,m}^{(r,s)}(A)=\sum_{n\in\Lambda_j}
\exp{ [{-n\pi \over 2\tau_2}(r+s{\bar\tau})a
+{n\over m+2}(r+s\tau){\partial\over\partial a}]}$.

Consider now the quantization of the $B$ field in \treintaytres. The level one
case ($m=1$) is somehow special. It can be easily shown that  for $m=1$ (and
$a\rightarrow b$) we can absorb in \treintaycuatro\ the $\Pi$ factors of the
denominator and write the character in the form
\REF\FK{I.B. Frenkel and V.G. Kac, {\sl Invent. Math.} {\bf 62} (1980) 301}
\REF\Segal{G. Segal \journal\cmp&80(81)301}
[\FK ,\Segal],
$$
\Phi_{l,1}(b)={\ex^{\pi b^2\over 4\tau_2}
\theta \Big [{{l\over 2}\atop 0}\Big ](b|2\tau)
\over \eta(\tau)},
\,\,\,\,\,\,\,\,\,\,\,\,\,\,
l=0,1.
\eqn\treintayseis
$$
Therefore, in this case, the $SU(2)_1$ characters can be written as $U(1)_1$
characters. Accordingly, the Verlinde operators that act correctly on the
states
\treintayseis\ can be represented by abelian Wilson lines
$W_{l,1}^\gamma(B)=\exp [-l\int_\gamma B]$ (notice that there is no trace!)
with  $B={\pi b\over 2\tau_2}\bar\omega- {\pi\bar b\over 2\tau_2}\omega$
and now we have the basic commutator $[\bar b,b]={2\tau_2\over
\pi}$ (which can be obtained by looking at the measure in the space of
level-one abelian characters).

Using the above results we can try to follow the spirit of the
abelian coset construction for this $SU(2)$ case. First of all, in complete
agreement with \quince, let us define new variables $c$ and $d$ as
follows:
$$
c={(m+2)a+b\over m+3},
\,\,\,\,\,\,\,\,\,\,\,\,\,
d={a-b\over m+3}.
\eqn\treintaysiete
$$
Notice that using the commutation relations for the $a$ and $b$
variables one gets $[\bar c,c]={2\tau_2\over\pi (m+3)}$, which
suggests that $c$ is a good variable for  $SU(2)_{m+1}$ wave
functionals. We have learned from our abelian example that, in
order to extract $SU(2)_{m+1}$ wave functions from $SU(2)_m\times
SU(2)_1$ states, one has to introduce a projector. In the non-abelian
case the Weyl group plays a major role. In the basis we have chosen to
describe our states (the $\lambda$'s in \treintaycuatro) the wave
functions are Weyl antisymmetric. As we argued above $c$ is the natural
variable for $SU(2)_{m+1}$ characters. It is thus clear that it will only be
possible to decompose into  $SU(2)_{m+1}$ states that sector of the
$SU(2)_m\times SU(2)_1$ Hilbert space which is antisymmetric under
$c\rightarrow -c$. Let us call $\cal P$  the operator that acting on any
function $f(c)$ gives $(f(c)-f(-c))/2$. Following steps similar to those that
led us to \veinte\ we get,
$$
{\cal P} (\lambda_{j,m+2}(a)\Phi_{l,1}(b))=
\sum_{i=0}^{m+1}\lambda_{i,m+3}(c)\chi_{j+1,i+1}(d),
\eqn\treintayocho
$$
where the $\lambda_{i,m+3}(c)$ are given by \treintaycinco\ with $m\rightarrow
m+1$ and $a\rightarrow c$. The sum is performed over $j-i$ even (odd) for
$l=0$ $(l=1)$. The functions $\chi_{p,q}(d)$ are
$$
\chi_{p,q}(d,\tau)={\ex^{{\pi k d^2\over 4\tau_2}}\over 2\eta(\tau)}
\Big [\Theta_{n_-,k}(d,\tau,0)-\Theta_{n_+,k}(d,\tau,0)
+\Theta_{-n_-,k}(d,\tau,0)-\Theta_{-n_+,k}(d,\tau,0)\Big],
\eqn\treintaynueve
$$
where $n_{\pm}=p(m+3)\pm q(m+2)$, and $k=(m+2)(m+3)$. The wave functions
\treintaynueve\ evaluated at the origin ($d=0$) are the familiar Rocha-Caridi
\REF\Rocha{A. Rocha-Caridi, in  {\it Vextex Operators in Mathematics and
Physics}, ed. by J. Lepowsky, S. Mandelstam and I. Singer, MSRI Publication \#
3, Springer, Heidelberg, 1984, 451-473}[\Rocha] characters for the minimal
unitary models with central charge $c_m=1-{6\over (m+2)(m+3)}<1$. Notice that
at
$d=0$ the last two terms of \treintaynueve\ equal the first two ones. From
\treintayocho\ we see that the range of $p$ ($q$) in $\chi_{p,q}$ is $1\leq
p\leq m+1$ ($1\leq q\leq m+2$). These values define the so-called conformal
grid. On the other hand, different values of $p,q$ in the conformal grid give
rise to the same character $\chi_{p,q}$. In fact, if we change $p\rightarrow
m+2-p$ and  $q\rightarrow m+3-q$, the $n_{\pm}$ change as $n_+\rightarrow
-n_++2k$, $n_-\rightarrow -n_-$. From our expression \treintaynueve\ we get
$\chi_{p,q}=\chi_{m+2-p,m+3-q}$. This is the well-known reflection property of
the conformal grid [\Gins]. Notice that at $d \not= 0$ it is essential to have
the four terms in \treintaynueve\ in order to fulfill this property. Finally,
we
can check that our wave functions $\chi_{p,q}$ behave under the modular group
exactly as the characters of the minimal models,
$$
\eqalign{
\chi_{p,q}(d,\tau)_{|_{T}}\equiv \chi_{p,q}(d,\tau+1)=
\ex^{2\pi i (h_{p,q}-{c_m\over 24})}\chi_{p,q}(d,\tau),\cr
\chi_{p,q}(d,\tau)_{|_{S}}\equiv \chi_{p,q}({d\over \tau},{-1\over\tau})=
\sum_{p',q'} S_{p,q}^{p',q'}\chi_{p',q'}(d,\tau),\cr}
\eqn\cuarenta
$$
where
$h_{p,q}={[(m+3)p-(m+2)q]^2-1\over 4(m+2)(m+3)}$
are the conformal weights of the primary fields and
$$
S_{p,q}^{p',q'}=\big ( {8\over (m+2)(m+3)}\big )^{{1\over2}}
(-1)^{(p+q)(p'+q')}\sin{\pi p p'\over m+2}\sin{\pi q q'\over m+3}.
\eqn\cuarentayuno
$$
The above results imply that the states of the effective problem in $d$ (\ie,
of the ${SU(2)_m\times SU(2)_1\over SU(2)_{m+1}}$ CS coset theory) are in
one-to-one correspondence with the characters of the $c_m<1$ unitary minimal
models. We have thus reproduced the GKO construction of these models within the
CS context.

In order to complete our effective quantum-mechanical problem for the coset
model we have to find the observables that act on the space of characters
\treintaynueve. These observables must be represented by topologically
invariant operators. The obvious choice are Wilson line operators associated
to homotopy classes of curves on $T^2$. First of all we must define a one-form
gauge connection for the variable $d$. In complete analogy with what we have
done for the other variables, let us define the abelian one-form
$D={\pi d\over 2\tau_2}\bar\omega- {\pi\bar d\over 2\tau_2}\omega$. From
the definition \treintaysiete\ and the commutation relations of $a$ and $b$
we get that
$[\bar d,d]={2\tau_2\over \pi (m+2)(m+3)}$, which implies the operator
representation $\bar d={2\tau_2\over \pi (m+2)(m+3)}{\partial\over\partial d}$.
For a given closed curve we will associate observables constructed by combining
abelian Wilson line operators for the gauge field $D$ with different charges.
We do not want  these operators  to take us out of the Hilbert space
spanned by the states \treintaynueve. Therefore we must choose the charges
entering into the combination  in such a way that the
resulting operator carries the quantum numbers of the characters $\chi_{p,q}$.
In this way we will get an association between states and operators which is
to be expected for a topological gauge theory. A glance at equation
\treintayocho\ reveals the origin of the quantum numbers $p$ and $q$ in
$\chi_{p,q}$: ${p-1\over2}$ is the $SU(2)_m$ isospin of the initial state
(before any projection) whereas ${q-1\over2}$  is the  isospin of the
$SU(2)_{m+1}$ state that we mod out when we perform the coset construction.
Therefore it is natural to suspect that the $U(1)$ charges needed to build up
Virasoro quantum numbers can be obtained by combining the weights of two
$SU(2)$ representations of isospins ${p-1\over2}$ and ${q-1\over2}$. Let us
check that this is indeed the case. Define the set of weights
$\Gamma_{p,q}=(m+3)\Lambda_{p-1}+(m+2)\Lambda_{q-1}$ and the operators
$W_{p,q}^{\gamma}=\sum_{n\in \Gamma_{p,q}}\ex^{-n\int_{\gamma}D}$. For a toral
knot $\gamma=r{\bf A}+s{\bf B}$ we can write the explicit representation of
$W_{p,q}^{(r,s)}$:
$$
W_{p,q}^{(r,s)}=\sum_{n\in \Gamma_{p,q}}
\exp [-{n\pi\over 2\tau_2}(r+s\bar\tau)d +{n\over (m+2)(m+3)}(r+s\tau)
{\partial \over \partial d}].
\eqn\cuarentaydos
$$
After some calculations we obtain the result of acting with the operator
\cuarentaydos\ on an arbitrary state,
$$
W_{p,q}^{(r,s)}\chi_{p',q'}(d,\tau)=
\sum_{n\in \Lambda_{p-1}\,\, l\in\Lambda_{q-1}}
M_{nl}^{(r,s)}(p,q;p',q')\chi_{p'+sn,q'+sl}(d,\tau),
\eqn\cuarentaytres
$$
where,
$$
M_{nl}^{(r,s)}(p,q;p',q')=
\ex^{i\pi r[s(p-1)(q-1)+(p-1)q'+(q-1)p'+{m+3\over m+2}({sn^2\over 2}+p'n)+
{m+2\over m+3}({sl^2\over2}+q'l)]}.
\eqn\cuarentaycuatro
$$
{}From eq. \cuarentaycuatro\ we confirm that the operators \cuarentaytres\ map
the Virasoro characters \treintaynueve\ into themselves. In particular from the
general expressions given above it is easy to check that the operators
$W_{p,q}^{(0,1)}$ associated to the non-contractible ${\bf B}$-cycle create the
state $\chi_{p,q}$ when acting on the vacuum $\chi_{1,1}$,
$$
W_{p,q}^{(0,1)}\chi_{1,1}=\chi_{p,q}.
\eqn\cuarentaycinco
$$
This last equation show us that the operators \cuarentaydos\ really carry the
quantum numbers associated to the degenerate representations of the Virasoro
algebra. As expected the operators associated to the contractible
${\bf A}$-cycle act diagonally on the space of characters. Their matrix
elements
are ratios of elements of the modular $S$ matrix,
$$
W_{p,q}^{(1,0)}\chi_{p',q'}=
{S_{p,q}^{p'q'}\over S_{1,1}^{p'q'}}\chi_{p'q'}.
\eqn\cuarentayseis
$$
Equations \cuarentaycinco\ and \cuarentayseis\ imply that
$W_{p,q}^{(r,s)}$ are
the Verlinde operators [\verope] associated to  arbitrary toral knots.
Performin
   g
modular transformations we can relate operators having different values of $r$
and $s$ in \cuarentaydos. In particular, one can get the ${\bf B}$-cycle
operators by conjugating the operators corresponding to the ${\bf A}$-cycle
with the $S$ matrix \cuarentayuno. This fact, together with the
Verlinde theorem [\verope], ensures that, for a fixed $(r,s)$ toral knot, the
operators \cuarentaytres\ satisfy the fusion rules of the corresponding primary
fields of the 2D conformal field theories.

{\it 4. Concluding remarks}
In this work we have presented a brief account of our main results. There
are several possible applications of our approach. For example, from the
general matrix element \cuarentaytres\ we could compute knot polynomials
\REF\kauf{For a review see L. H. Kauffman, {\it Knots and Physics}, World
Scientific, Singapore, 1991, and references therein.}[\kauf] for
the unitary minimal Virasoro models in the same way as they were computed for
the WZW model in Ref. [\knot]. It is also possible in principle to extend our
formalism to other coset theories such as, for example, the $N=1$
supersymmetry minimal models (obtained in [\GKO] from the coset models
${SU(2)_m\times SU(2)_2\over SU(2)_{m+2}}$) and the minimal series of the $W_N$
algebras  (described by the coset models  ${SU(N)_m\times SU(N)_1\over
SU(N)_{m+1}}$). We expect to report on these issues in a future work.

\refout
\end